\documentclass[aps,twocolumn,groupedaddress]{revtex4}
\usepackage{amsmath,amssymb}

\begin{document}
\bibliographystyle{apsrev}


\title{Geometrical Brownian Motion Driven by  Color Noise}




\author{Ryszard Zygad\l{}o }
\affiliation{Marian Smoluchowski
Institute of Physics, Jagiellonian University, Reymonta 4, PL--30059 Krak\'ow,
Poland}


\date{\today}

\begin{abstract}
The evolution of prices on ideal market is given by geometrical Brownian motion,
where Gaussian white noise describes fluctuations. We study the effect
of correlations introduced by a color noise.
\end{abstract}

\maketitle


\section{Introduction}
The continuous--time description of financial markets frequently
uses the (multiplicative, driftless) Ito stochastic differential
equations (SDE). The market is considered to be fair in a sense
that, independently on the strategy, the expected future value of an
investment is equal, if discounted prices are used, to the value of
the capital used. An attempt to make the price movements  
partially predictable is to replace the driving white noise by a
correlated process. Because the direct white noise limit procedure
corresponds to the Stratonovich interpretation of the resulting SDE,
the initial equation should be first properly rewritten. The paper
is organized as follows. In Sec.~2 some basic results of SDE theory
are collected, in Sec.~3 the Black and Scholes equation  is
introduced.  Two examples of color noises are presented in Sec.~4,
which are then used, in Secs.~5 and 6, to generalize the Black and
Scholes model.   The last section contains a short summary and
conclusions.

\section{SDE. Ito vs Stratonovich}
The usual assumption concerning ideal market is that the evolution of
appropriately {\it discounted prices} represents a {\it fair game}, i.e.,
that it is described by a certain {\it martingale}.
Such processes, by the definition,
have  the property that the (future) conditional  mean value
$\langle x_t| x_{t_n},\ldots, x_{t_1} \rangle = x_{t_n}$,
$t\ge t_n \ge \ldots \ge t_1$, is equal to the last value specified by
the condition [1,2]. An important class of martingales are
{\it driftless} Ito processes [1-4]
\begin{equation}
 dx_t = f(x)dt + g(x) d\circ W_t,
\end{equation}
where $f(x)=0$ and
the $\circ$ sign is to indicate that the equation is
{\it interpreted} according to the Ito definition that
$\langle g(x) d\circ W_t \rangle = 0$
({\it nonanticipating} property). $W_t$ is the Wiener process
normalized by the condition
\begin{equation}
\langle \exp(y W_t) \rangle = \exp(Dt y^2),     \end{equation}
or, equivalently,
\begin{equation}
\langle \xi_t \xi_0\rangle = 2D\delta(t),     \end{equation}
where $\xi_t \equiv dW_t/dt $ is a Gaussian white noise (GWN).

The well known consequence  of the nonanticipating property is
that the ordinary rules of differentiation and integration are no longer
valid [1-4], being replaced by the specific Ito calculus. Particularly,
it turns out that  $x_t \equiv x(t, W_t)$,
considered as a function of two variables, represents the solution
to the Ito
Eq.~(1) only if the  usual condition $\partial x/ \partial W = g(x)$
and the {\it unusual} one
$\partial x/ \partial t =  f(x) - D g(x) g^\prime (x)$  are
satisfied. This means that using the ordinary calculus
the same process $x(t, W_t)$ is considered to be the solution
of the Stratonovich equation [2,4,5]
\begin{equation}
dx_t = [f(x)- Dg(x)g^\prime(x)]dt + g(x) d W_t
\end{equation}
(in our notation without $\circ$  sign)
and in such sense both  Eqs.~(1) and (4) are
equivalent.
The term $Dgg^\prime$ is called ``spurious drift.''
Note that the Stratonovich interpretation is more popular in a physical
literature because
the
well recognized (ordinary) methods of transforming the variables and
solving the differential equations can be used.

\section{Black and Scholes model}
In his pioneering work [6] Bachelier adopts {\it arithmetical} Brownian
motion
\begin{equation}
\dot{x}_t = rx + \xi_t
\end{equation}
to describe the evolution of stock prices. Here $r>0$ is an
intrinsic growth rate often identified simply with the
{\it interest} rate. Except for the sign $r = - \gamma <0$, where $\gamma$ is
the {\it friction} coefficient, it is
the famous Langevin equation, for the
Brownian particle's velocity, of  the Einstein--Smoluchowski
theory of Brownian motion [7]. Because the solutions of Eq.~(5) are Gaussian
(Ornstein--Uhlenbeck processes [8]), they are in fact not well
suited for modeling prices, which are the nonnegative quantities.

Assuming an independent and Gaussian character of the {\it relative} changes
one obtains the (Samuelson) Black and Scholes equation (B\&S) [9-11]
\begin{equation}
dS_t = rS dt +  S d\circ W_t
\end{equation}
or, equivalently,
\begin{equation}
\dot{S}_t = [r- D]S + S \xi_t
\end{equation}
if the Stratonovich interpretation is used. The stochastic solution
\begin{equation}
S_t = S_0 e^{(r-D)t} \exp(W_t)
\end{equation}
immediately follows from Eq.~(7). Using Eq.~(2) one verifies
that
\begin{equation}
\langle S_t \rangle = S_0 e^{(r-D)t} \langle \exp(W_t) \rangle = S_0 e^{rt}
\end{equation}
in agreement with Eq.~(6).
Eq.~(9) shows  that the average return, related to the passive
investment ``buy and hold,'' is determined by the
interest rate $r$. Because the  discounted price
$\tilde{S}_t = e^{-rt} S_t$ is a martingale  the consequence of
the games theory is that no other strategy
can lead to a better result.

\section{Color noise}
The Langevin random force (GWN) may be considered as a ``singular''
limit of certain ``regular'' correlated processes. Two important
examples are the following.

The Gaussian color noise
(GCN) (or the stationary Ornstein--Uhlenbeck process) is defined as
the Gaussian process of a zero mean and an exponentially decaying
autocorrelation function [4,8]
\begin{equation}
K(t) \equiv \langle \xi_t \xi_0 \rangle = D\tau^{-1} \exp(-t/\tau),
\end{equation}
where $\tau >0$ is the correlation-time.
Because (as a generalized function)
$K(t) \to 2D\delta (t)$ for $\tau \to 0$ the GCN (10)
approaches GWN (3) if the correlation-time goes to zero.

Another exponentially correlated process, of a different origin, is
the dichotomous Markov process (DM) $\xi_t = \sigma (-1)^{N_t}$,
where $\sigma$ is a binary variable equal $\pm |\sigma|$ with
probability $1/2$ and $N_t$ is a Poisson counting process with
parameter $\lambda$ [12-14],
\begin{equation}
 \langle \xi_t \xi_0 \rangle = \sigma^2 \exp(-2\lambda t).
\end{equation}
It may be shown that at
the limit $\sigma^2 \to \infty$, $\lambda \to \infty$,
$\sigma^2/2\lambda = D = {\rm const}$ the GWN (3) is also
recovered [12]. The correlation-time of DM is $1/2\lambda$ $(= \tau)$.

Note that the above mentioned GWN-limit procedures are consistent
with the Stratonovich interpretation. Thus we will study
Eq.~(7) with color noise (10) or (11).
Without loss
of generality we assume that $r=0$, which corresponds to
the use of discounted prices.

\section{B\&S model with GCN}
Let $r=0$ and
\begin{equation}
\dot{S}_t = - DS + S \xi_t,
\end{equation}
where $\xi_t$ is GCN (10).
Then
\begin{equation}
S_t = S_0 e^{-Dt} \exp \Bigl[ \int_0^t \xi_s ds \Bigr].
\end{equation}
Using the general formula for stationary Gaussian processes and
after that
Eq.~(10)
\begin{eqnarray}
\Bigl\langle \exp \Bigl[ \int_0^t \xi_s ds \Bigr] \Bigr\rangle && =
 \exp \Bigl[ \int_0^t ds_1 \int_0^{s_1}  ds_2 K(s_1-s_2) \Bigr]  \nonumber \\
&&= \exp \Bigl[ Dt - D\tau(1-e^{-t/\tau}) \Bigr], 
\end{eqnarray}
one obtains
\begin{equation}
\langle S_t \rangle =
S_0  \exp \Bigl[-D\tau(1-e^{-t/\tau}) \Bigr] \approx S_0 e^{-D\tau}.
\end{equation}
The result (15) may be summarized as follows.  The presence of a
color noise in Eq.~(12) introduces certain correlations between
successive changes of prices. In contrast to the ideal market model
the historical information about prices can be in principle useful
to improve the strategy of investment. The cost to be payed is that
the passive long-time investment leads to the partial loss of the
initial capital $S_0$, by a constant factor $e^{-D\tau}$ $(\approx
1-D\tau)$ in discounted prices.

\section{B\&S model with DM}
Assume again that $r=0$ and consider
\begin{equation}
\dot{S}_t = - (\sigma^2/2\lambda)S + S \xi_t,
\end{equation}
where $\xi_t$ is the asynchronic binary noise (11). Because
\begin{equation}
S_t = S_0 e^{-\sigma^2 t/2\lambda} \exp \Bigl[ \int_0^t \xi_s ds \Bigr]
\end{equation}
we need
\begin{equation}
\Phi(t) = \Bigl\langle \exp \Bigl[ \int_0^t \xi_s ds \Bigr] \Bigr\rangle
\end{equation}
in order to compute certain averages. Let
\begin{equation}
\Psi(t) = \Bigl\langle \xi_t \exp \Bigl[ \int_0^t \xi_s ds \Bigr] \Bigr\rangle.
\end{equation}
Then $\dot\Phi = \Psi$ and $\dot\Psi = - 2\lambda \Psi + \sigma^2 \Phi$, where
the latter equation follows from Shapiro--Loginov formula [15].
The solution of
\begin{equation}
\ddot\Phi + 2\lambda \dot\Phi - \sigma^2 \Phi =0
\end{equation}
satisfying $\Phi(0)=1$, ${\dot\Phi}(0)=0$ is
\begin{equation}
\Phi(t) = \Bigl({1\over 2} +{1\over 2q}\Bigr) e^{\lambda(q-1)t} +\Bigl(
{1\over 2} -{1\over 2q}\Bigr) e^{-\lambda(q+1)t},
\end{equation}
where $q=\sqrt{1+ \sigma^2/\lambda^2}$.
For sufficiently long time and $(2D/\lambda =)$ $\sigma^2/\lambda^2 \ll 1$
\begin{equation}
\Phi(t) \approx \Bigl( 1 - {1\over 4} {\sigma^2\over \lambda^2} \Bigr)
\exp \Bigl( {\sigma^2 t \over 2 \lambda} - {\sigma^4 t \over 8 \lambda^3} + \ldots
\Bigr)
\end{equation}
and thus
\begin{equation}
\langle S_t \rangle \approx
S_0  (1-D\tau)  e^{-D^2\tau t},
\end{equation}
where $\tau = 1/2\lambda$ and $D=\sigma^2/2\lambda$. Comparing to
Eq.~(15) the case (23) seems even worse generating the still
increasing (with time) losses in discounted prices (or decreasing
the intrinsic growth rate from $r$ to $r-D^2\tau$ in real prices).
On the other hand the ``buy and hold'' strategy, for selfevident
reasons, is quite inappropriate for this case. In fact Eq.~(17)
shows that the {\it realization} of $S_t$ consists of the periods of
exponential decay  $S \sim e^{-(|\sigma|+D)\Delta t}$ separated by
the periods of exponential growth $S \sim e^{(|\sigma|-D)\Delta t}$
(if $|\sigma| < 2\lambda$; otherwise the price always falls). The
length of the periods is random with the average equal to
$1/\lambda$. Moreover the trajectory of $S_t$ is continuous.
``Playing with trend'' one buys the stock at the beginning of a
growth period and sells immediately when the move changes the
direction. The distribution of waiting times  for DM is given by
$p(t)=\lambda e^{-\lambda t}$ and the corresponding price $S(t)=S_0
e^{(|\sigma|-\sigma^2/2\lambda)t}$, so the expected return per one
cycle of an investment is
\begin{equation}
\bar{S}= \int_0^\infty p(t) S(t) dt = S_0 {2 \over
(1-|\sigma|/\lambda)^2 + 1}.
\end{equation}
Note that at the GWN-limit, $|\sigma|=\sqrt{2\lambda D}$,
$\lambda \to \infty$, the r.h.s.\ of Eq.~(24) is (still) equal $S_0$,
which reflects the {\it fairness} of the ideal market. The ratio
$\bar{S}/S_0 >1 $ if $|\sigma|/\lambda <2$ (or $D<2\lambda$ ). The
maximum $\bar{S}/S_0 =2$ corresponds to $|\sigma|=\lambda = 2D$.
Thus, in spite of the general decreasing tendency (23), the market
described by Eq.~(16) provides easy earn opportunities.

\section{Remarks}
The B\&S equation, written in the Stratonovich form (7), can be
easily generalized by an appropriate replacement of the driving
noise. The B\&S equation with a color noise remains exactly
solvable. The particular cases of GCN and DM have been analyzed in
Secs.~5 and 6. The general conclusion is the following. The limit of
zero correlation--time corresponds to the ordinary B\&S model, when
the discounted price $\tilde{S}_t = e^{-rt} S_t$ is a martingale and
the expected future price is $\langle S_t \rangle = S_0 e^{rt}$. On
the correlated market ($\tau >0$) the expected price is lowered:
$\langle S_t \rangle \approx S_0 e^{-D\tau} e^{rt} \approx S_0
(1-D\tau) e^{rt}$, Eq.~(15), for GCN and $\langle S_t \rangle
\approx S_0 (1-D\tau) e^{(r-D^2\tau)t}$, Eq.~(23), for DM,
respectively. Thus the long--time investment is not particularly
recommended. On the other hand, in the presence of correlations  the
historical prices contain a certain information which can be used to
improve the investment. In the case of DM it is easy to identify and
use the short--time trends to get a certain earn, as shows Eq.~(24).


\begin{thebibliography}{}

\bibitem{1}
Y.V. Prokhorov and Y.A. Rozanov, {\em Probability Theory}
(Springer, Berlin, 1969);
I.I.~Gihman and A.V.~Skorohod, {\em Stochastic Differential Equations}
(Springer, Berlin, 1972).

\bibitem{2}
R. Zygad\l{}o, Phys.\ Rev.\ E {\bf 68},  046117 (2003).

\bibitem{3}
K.~Ito, Mem.\ Am.\ Math.\ Soc.\ {\bf 4}, 289 (1951).

\bibitem{4}
C.W.\ Gardiner,
{\em Handbook of Stochastic Methods} (Springer, Berlin, 1983).

\bibitem{5}
R.L. Stratonovich, {\em Topics in the Theory of Random Noise}
(Gordon and Breach, New York, 1967).


\bibitem{6}
L. Bachelier, Ann.\ Sci.\ Ecol.\ Norm.\ Sop.\ {\bf 17}, 21 (1900).

\bibitem{7}
A.~Einstein, {Ann.\ Physik} {\bf 17}, 549 (1905);
M.~Smoluchowski, {Ann.\ Physik} {\bf 21}, 756 (1906);
P.~Langevin, {Comptes rendus} {\bf 146}, 530 (1908).

\bibitem{8}
G.E.~Uhlenbeck, L.S.~Ornstein, {Phys.\ Rev.\ }{\bf 36}, 823 (1930).


\bibitem{9}
P.A. Samuelson, Industrial Management Review {\bf 6}, 13 (1965).

\bibitem{10}
F. Black and M. Scholes, J.\ Polit.\ Economy {\bf 81}, 637 (1973).

\bibitem{11}
R.N. Mantegna and H.E. Stanley
{\em Introduction to Econophysics: Correlations and Complexity in Finance}
(University, Cambridge, 1999).

\bibitem{12}
C. van den Broeck, J.\ Stat.\ Phys.\ {\bf 31}, 447 (1983).

\bibitem{13}
W.~Horsthemke and R.~Lefever,
{\em Noise-Induced Transitions} (Springer, Berlin, 1984).

\bibitem{14}
R. Zygad\l{}o, Phys.\ Rev.\ E {\bf 54}, 5964 (1996).


\bibitem{15}
V.E.~Shapiro and V.M.~Loginov, Physica A {\bf 91}, 563 (1978);
V.M.~Loginov, Acta Phys.\ Pol.\ B {\bf 27}, 693, (1996).


\end{thebibliography}
\end{document}